\documentclass[12pt,a4paper]{article}
\usepackage[utf8]{inputenc}
\usepackage[english]{babel}

\usepackage{amsmath}
\usepackage{amsfonts}
\usepackage{makeidx}

\usepackage{hyperref}
\hypersetup{colorlinks=true,linkcolor=blue,filecolor=magenta,urlcolor=cyan}

\usepackage[style=numeric,sorting=ynt,citestyle=numeric-comp]{biblatex}
\bibliography{References}

\usepackage{float}
\usepackage{subcaption}

\usepackage{csquotes}
\usepackage{amssymb}
\usepackage{graphicx}
\usepackage{kpfonts}
\usepackage[left=2cm,right=2cm,top=2cm,bottom=2cm]{geometry}
\title{SARS-COV-2 Pandemic: Understanding the Impact of Lockdown in the Most Affected States of India}
\author{Chinmay Patwardhan\\ 
\emph{\small Department of Mathematics,}\\
\emph{\small Institute of Chemical Technology, Mumbai, India}\\
\emph{\small e-mail: chinmayp06@gmail.com}}
\date{April 27, 2020}

\begin{document}
\maketitle
\begin{abstract}
	SARS-COV-2 has stopped the world in its footsteps and a third of the population has been forced to stay at home. Here we present a comparative study of the performance of states of India, in curbing the spread of the disease, that are most affected by the pandemic. We have done so based on the data collected between 14$ ^{th} $ March, 2020 to 17$ ^{th} $ April, 2020. We do this by comparing the smoothened time series and percentage changes along with change point detection of the daily confirmed cases. We also discuss the different policies and strategies adopted by the states in curbing the disease and the ground level implementation. We have developed ARIMA(p,d,q) models where (p,d,q) are obtained by minimising the AIK(Akaike Information Criterion). These models are used to make forecasts for the country which show that by 7$ ^{th} $ May, 2020, India would have around 50,237 confirmed cases and a doubling rate of 11 days. On the basis of the performance, of each state, we argue that a local level strategy which is based on the demographics of that particular region be developed instead of a central and uniform one. 
\end{abstract}
\section{Introduction}
The SARS-COV-2(Severe Acute Respiratory Syndrome Coronavirus 2)  outbreak occured, in Hubei province of China, in December 2019 and has within a few months of the outbreak grown into a pandemic\cite{WHOevent}. As of 27$^{th}$ April, 2020, over 2,810,325 people have been infected with the disease and 193,825 have deceased\cite{WHO} in over 213 countries. Such rapid spread of the virus has mainly been successful due to a lack of preventive measures in the initial days of the outbreak, vast volume of international travel and its mode of infection i.e. through personal contact and aerosols\cite{WHOtransmission}. In order to contain the virus China was the first country to impose a full lock down of the epicentre of the virus, i.e. Hubei province, on the 23$^{rd}$ of March 2020\cite{ChinaLock}. Subsequently a similar strategy was adopted in Italy \cite{ItalyLock} and later on in several parts of the world. \newline

The first case in India was detected on 30$^{th}$ Jan, 2020\cite{IND1st} in Kerala and since then a total of 27,982 confirmed cases, with 6523 recovered and 884 deaths (as of Thursday, 27$ ^{th} $ April 2020), have been reported. The ministry of health and family welfare, responsible for controlling the disease, has maintained its stand that the epidemic is in the second stage i.e. local transmission. The socio-economic impact of community transmission, of SARS-COV-2 in India, could be devastating, \textbf{Jha et al.} in their paper\cite{commtrans},discussed the possibility of community transmission in India and its capability to tackle an uncontrolled outbreak. They argued in the light of availability of resources such as regular and ICU beds, PPE (Personal Protective Equipment) for health care workers and the number of doctors, and suggest three possible methods to curb the spread, which are isolation of the infected, quarantine and social distancing. With a population of ~1.3 billion people it is a real challenge to identify and isolate individuals in case of community transmission. Several measures were taken by the Govt. of India in order to curb and control the spread of SARS-COV-2. Initially passengers, coming from highly affected countries like China, South Korea, Germany and Italy, were screened and  home quarantined and later all flights were grounded and international travel halted\cite{airban,MOHFW}. The Govt. of India declared a country-wide lock down on the 25$^{th}$ of March,2020\cite{Ind}, while some states like Andhra Pradesh, Kerala, Delhi, Maharashtra and Assam did so earlier on the 23$^{rd}$ of March, 2020 \cite{AP,MH,KL,DL}. \newline 
 
As a result of the lock down it is expected that the rate of growth reduces after the lock down and the number of daily cases start declining; this sort of a trend has been observed in several countries like China\cite{peng2020epidemic}, Italy, Spain etc.\citeauthor{outbreakanal} \cite{outbreakanal}, have done a study on the effectiveness of the 21 days national lock down in India and made predictions using exponential and polynomial regression models. A nation-wide study is very helpful for the central government to implement policies but a further study is required to device a state-wise strategy for dealing with COVID-19 (Coronavirus Disease 19) pandemic. As it is obvious from the data that not every state in the country is affected equally and in a similar manner, for instance, Maharashtra has over 5000 confirmed cases (as of 1012 hrs, Thursday, 23 April 2020 (IST)) \cite{datasource} whereas Jharkhand, Tripura, Mizoram or Goa have below 50, and hence it is necessary to study the spread of the virus in each of these most affected states separately.  \newline

In our study, instead of looking at the country as a whole, we focus on the 14 states and a union territory of India which are most affected by the SARS-COV-2 pandemic. These are Andhra Pradesh, Delhi, Gujarat, Haryana, Jammu and Kashmir, Karnataka, Kerala, Madhya Pradesh, Maharashtra, Punjab, Rajasthan, Tamil Nadu, Telangana, Uttar Pradesh and West Bengal. In this study we have dealt with the following questions,
\begin{enumerate}
	\item Have the number of the daily cases, in each state under consideration, reduced as a result of lock down?
	\item  When do we see a peak in the number of cases and can it be attributed to the lock down?
	\item  How do different states compare to each other in the effectiveness of the lock down and the response to tackling COVID-19?
	\item  Make short term predictions based on the already available data
\end{enumerate}

\section{Methods}
\subsection{Data Source}
We have used 35 data points beginning from 14$^{th}$ March, 2020 till 17$^{th}$ April, 2020, which cover a small period prior to the lock down and the entirety of the lockdown period. This data has been obtained from a crowd sourced patient database \cite{datasource}

\subsection{Methodology}

We concentrate on the number of daily confirmed cases instead of the cumulative number of cases as the variance in the data, for the cumulative cases, is higher than that in the daily confirmed cases. If we assume $ X_{k} $ to be a random variable denoting the number of confirmed cases on the $ k^{th} $ day, and $ Y_{k}  = \sum_{j = 1}^{k}X_{j}$, i.e. the cumulative number of cases on the $ k^{th} $. Then if $ Var(X_{k}) = \sigma_{k}^{2}  $ for each $ k $, then $ Var(Y_{k}) = \sum_{j = 1}^{k}Var(X_{j}) + \sum_{i \neq j = 1}^{k}Cov(X_{i},X_{j}) $, which is always greater than $ \sigma_{k} $   i.e. the variance of the random variable denoting the number of confirmed cases on the k$^{th}$ day. For this reason we focus on studying and modelling the number of daily confirmed cases instead of cumulative number of cases.\newline

We individually assess the rate of change in the daily number of confirmed cases for each state by calculating the percentage change. If $ x_{k} $ is the number of cases on the $ k^{th} $ day and $ x_{k-1} $ is the number of cases on the $ (k-1)^{th} $ day, then the percentage change on the $ k^{th} $ day is given by $ \frac{x_{k} - x_{k-1}}{x_{k-1}}\times 100 $. These percentage changes are plotted for each state and studied by isolating them into three components,(a) pre-lock down to start of lock down, (b)start of lock down to a week after lock down and (c) a week after lock down until the end of the lock down\label{1}. The purpose of doing so is to compare the percentage change in the number of daily cases in the different periods and see if there is any statistically significant change between post lock down and pre lock down periods.\newline
 
While the percent changes give a good picture as to whether the rate of change of the daily cases has a decreasing trend or an increasing trend, they do not tell exactly when the number of cases start decreasing or increasing. For this reason we used the \textit{Pandas} package in Python 3 to do a offline change point analysis of the time series data to identify four change-points. Change points in a time series are the points at which the time series shows abrupt changes and transition between different states of a time series. Change point detection is a very helpful tool while making predictions\cite{Aminikhanghahi2017}. These four change-points were studied with reference to the pre- and post- lock down periods mentioned above paragraph. Along with this, 3-point moving averages were plotted in order to find the peak in the daily number of cases. The purpose for selecting a 3-point moving averages scheme is to smooth the time series data while preserving the trend, reduce the impact of outliers and conserve some of the variation in the data. Compared to a 4-point which is far better at making the data smooth, it reduces the number of time series data points and also the averages are plotted at fictitious mid points. These peaks for each of the states were plotted and a cut-off based on the average incubation period (7-14 days) was used to determine whether the lock down has helped in flattening the curve. \newline
 
Finally, in order to make short term predictions, ARIMA (Auto-Regressive Integrated Moving Averages) model were fitted for each state. The data was checked for stationaryity using the ADF(Augmented Dicky Fuller) test\cite{ShumwayTSAbook} and the KPSS(Kwiatkowski-Phillips-Schmidt-Shin) test\cite{KPSStest}. Further, the selection of the number of auto-regressive variables ($ p $), the order of differencing ($ d $) and the number of moving averages variables ($ q $) was on the basis of the ACF(Auto-Correlation Function) and the PACF(Partial Auto-Correlation Function) and the AIC (Akaike Information Criterion). Based on the ACF plot the range of $ p $, say $ p_{r} $ was decided such that the correlation coefficient $ k^{th} $ lag is above 0.4 and similarly the range of $ q $, say$ q_{r} $, was decided using the PACF plot. Then for each $ p $,$ d $ and $ q $ value in the range $ (0,p_{r}) $, $ (0,q_{r}) $ and $ d = 0,1,2 $, the AIC value of the ARIMA model was computed and the tuple $ (p,d,q) $ with the least AIC value were selected. An ARIMA($ (p,d,q) $) was developed using the package \textit{statsmodels} and predictions were made on the basis of this model.

\section{State-wise Analysis}

\subsection{Kerala}
The Kerala state government imposed a complete lock down of the state, except for the movement of essential goods and associated services like hospitals, grocery shops etc., from 23$^{rd}$ March 2020\cite{KL}. \newline

\begin{figure}[H]
	\centering
	\begin{subfigure}[b]{0.3\textwidth}
		\includegraphics[width=\textwidth]{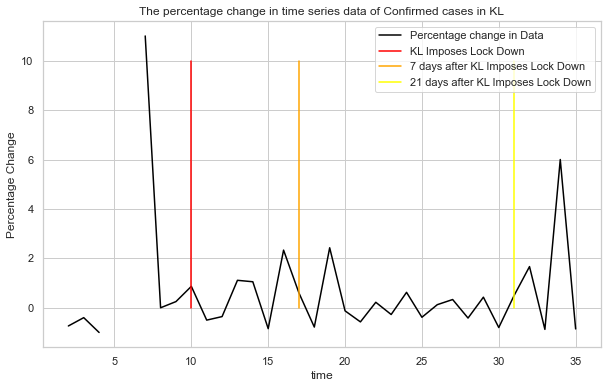}
		\label{fig:1a}
		\caption{}
	\end{subfigure}
	~ 
	\begin{subfigure}[b]{0.3\textwidth}
		\includegraphics[width=\textwidth]{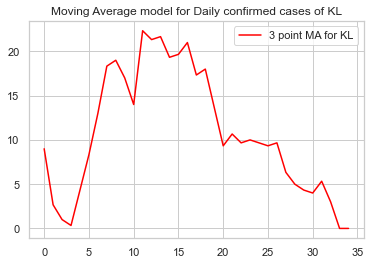}
		\label{fig:1b}
		\caption{}
	\end{subfigure}

	~ 
	\begin{subfigure}[b]{0.3\textwidth}
		\includegraphics[width=\textwidth]{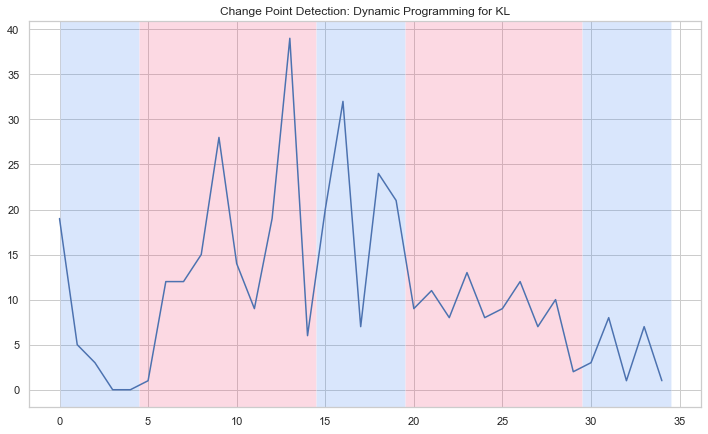}
		\label{fig:1c}
		\caption{}
	\end{subfigure}
	\begin{subfigure}[b]{0.3\textwidth}
		\includegraphics[width=\textwidth]{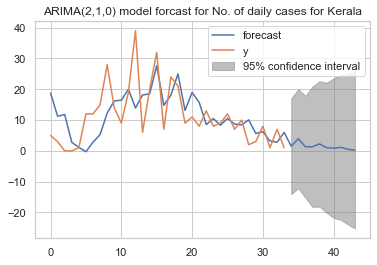}
		\label{fig:1d}
		\caption{}
	\end{subfigure}

	\caption{\small \sl (i)The figure (a) shows the percent change in the number of Daily confirmed cases of SARS-COV-2 in the state of Kerala along with vertical lines denoting the beginning of lock down, 7 days after the lock down and he end of lock down (ii) figure (b) is the plot of the 3-point moving averages for the time-series data of the state of Kerala (iii) figure (c) shows the change points in the the time series data for Kerala, the edges of the pink bands represent the change points and the bands with similar trends (iv) figure (d) is the ARIMA(2,1,0)  fitted model for the time-series data along with future predictions upto 1$^{st}$ May,2020 and the prediction interval. }\label{fig:1}
\end{figure}

\subsection{Maharashtra}
The Maharashtra state government imposed a complete lock down of the state, except for the movement of essential goods and associated services like hospitals, grocery shops etc., from 23$^{rd}$ March 2020\cite{MH}. \newline

\begin{figure}[H]
	\centering
	\begin{subfigure}[b]{0.3\textwidth}
		\includegraphics[width=\textwidth]{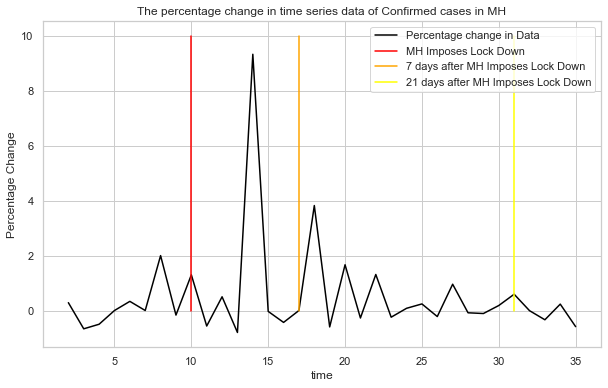}
		\label{fig:2a}
		\caption{}
	\end{subfigure}
	~ 
	\begin{subfigure}[b]{0.3\textwidth}
		\includegraphics[width=\textwidth]{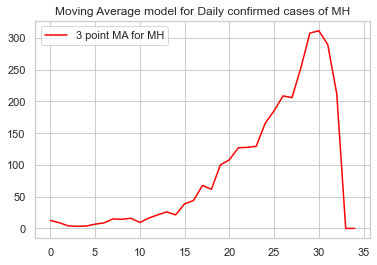}
		\label{fig:2b}
		\caption{}
	\end{subfigure}
	
	~ 
	\begin{subfigure}[b]{0.3\textwidth}
		\includegraphics[width=\textwidth]{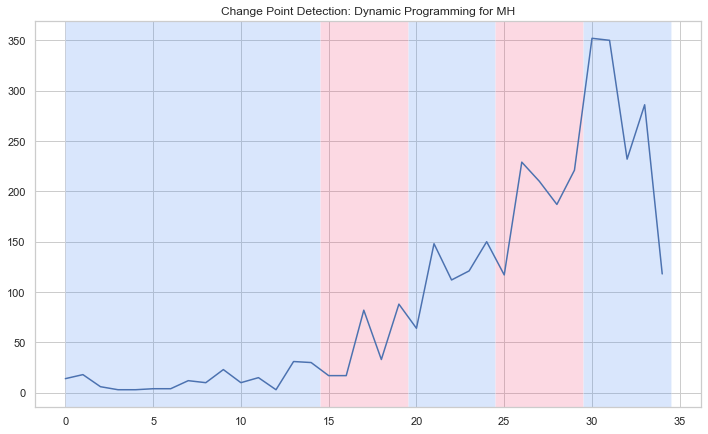}
		\label{fig:2c}
		\caption{}
	\end{subfigure}
	\begin{subfigure}[b]{0.3\textwidth}
		\includegraphics[width=\textwidth]{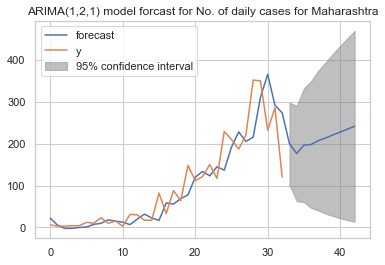}
		\label{fig:2d}
		\caption{}
	\end{subfigure}
	
	\caption{\small \sl (i)The figure (a) shows the percent change in the number of Daily confirmed cases of SARS-COV-2 in the state of Maharashtra along with vertical lines denoting the beginning of lock down, 7 days after the lock down and he end of lock down (ii) figure (b) is the plot of the 3-point moving averages for the time-series data of the state of Maharashtra (iii) figure (c) shows the change points in the the time series data for Maharashtra, the edges of the pink bands represent the change points and the bands with similar trends (iv) figure (d) is the ARIMA(1,2,1)  fitted model for the time-series data along with future predictions upto 1$^{st}$ May,2020 and the prediction interval. }\label{fig:2}
\end{figure}

\subsection{Punjab}
The Punjab state government imposed a complete lock down of the state, except for the movement of essential goods and associated services like hospitals, grocery shops etc., from 25$^{th}$ March 2020\cite{Ind} \newline
\begin{figure}[H]
	\centering
	\begin{subfigure}[b]{0.3\textwidth}
		\includegraphics[width=\textwidth]{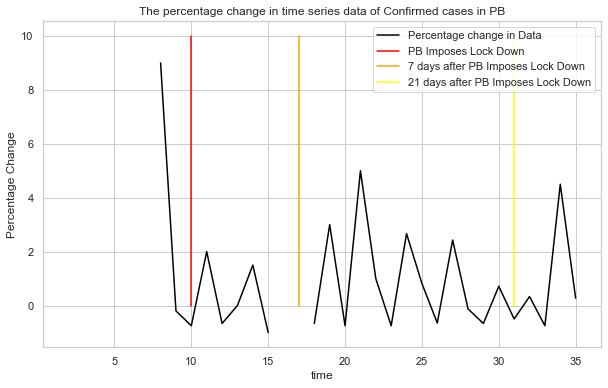}
		\label{fig:3a}
		\caption{}
	\end{subfigure}
	~ 
	\begin{subfigure}[b]{0.3\textwidth}
		\includegraphics[width=\textwidth]{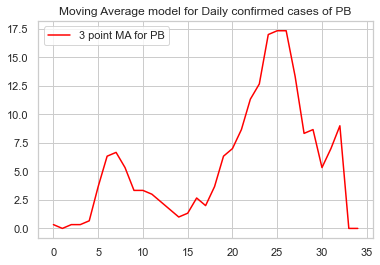}
		\label{fig:3b}
		\caption{}
	\end{subfigure}
	
	~ 
	\begin{subfigure}[b]{0.3\textwidth}
		\includegraphics[width=\textwidth]{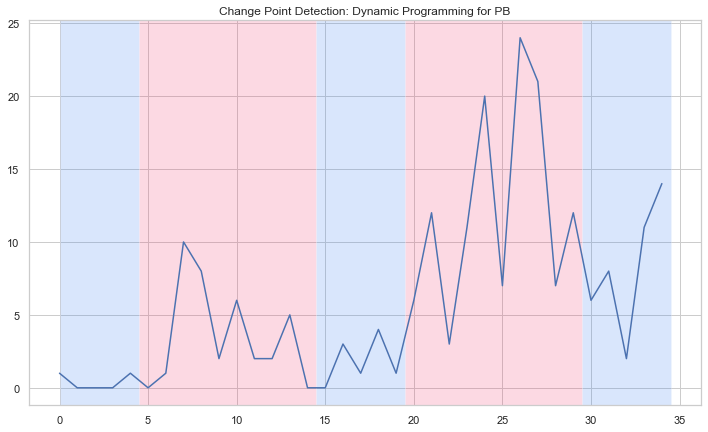}
		\label{fig:3c}
		\caption{}
	\end{subfigure}
	\begin{subfigure}[b]{0.3\textwidth}
		\includegraphics[width=\textwidth]{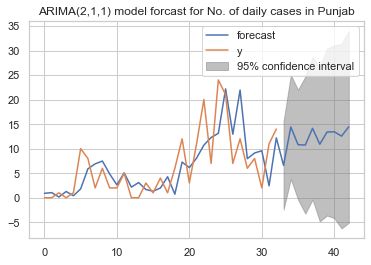}
		\label{fig:3d}
		\caption{}
	\end{subfigure}
	
	\caption{\small \sl (i)The figure (a) shows the percent change in the number of Daily confirmed cases of SARS-COV-2 in the state of Punjab along with vertical lines denoting the beginning of lock down, 7 days after the lock down and he end of lock down (ii) figure (b) is the plot of the 3-point moving averages for the time-series data of the state of Punjab (iii) figure (c) shows the change points in the the time series data for Punjab, the edges of the pink bands represent the change points and the bands with similar trends (iv) figure (d) is the ARIMA(2,1,1)  fitted model for the time-series data along with future predictions upto 1$^{st}$ May,2020 and the prediction interval. }\label{fig:3}
\end{figure}

\section{Discussion} 

\begin{table}[H]
	\centering
	\begin{tabular}{|l|l|}
		\hline
		States            & Peaked cases  \\ \hline
		Andhra Pradesh    & 04-Apr-20    \\
		Delhi             & 11-Apr-20    \\
		Gujarat           & 15-Apr-20    \\
		Haryana           & 06-Apr-20    \\
		Jammu and Kashmir\footnote{Union Territory of India} & 08-Apr-20    \\
		Karnataka         & 15-Apr-20    \\
		Kerala            & 25-Mar-20    \\
		Madhya Pradesh    & 15-Apr-20    \\
		Maharashtra       & 13-Apr-20    \\
		Punjab            & 08-Apr-20    \\
		Rajasthan         & 10-Apr-20    \\
		Tamil Nadu        & 01-Apr-20    \\
		Telangana         & 03-Apr-20    \\
		Uttar Pradesh     & 13-Apr-20    \\
		West Bengal       & 13-Apr-20   \\ \hline
	\end{tabular}
\label{Table:1}
\caption{\small \sl Table indicating the dates that the peak number of cases occur for each of the states based on the change point detection and 3-point moving averages graphs}
\end{table}

The Table:1 gives the dates at which each state experiences its peak number of cases. When we observe the dates at which the states have had their peak number of cases, some states like Andhra Pradesh, Haryana, Jammu and Kashmir, Tamil Nadu and Telangana have a peak within one week after implementation of the lock down whereas states like Gujarat, Maharashtra, Karnataka, Madhya Pradesh, Delhi attain their peak number of cases at around the end of the lock down period which suggests that these states are yet to attain their peak number of cases. This disparity among the states suggests that the implementation of the lock down may have been different in each state and can be attributed to other factors like a difference in policies and need further investigation. For example,  Maharashtra has the highest number of cases in the entire country which could be possibly attributed to the high air traffic, population density and the higher testing rate\cite{MHtestcap} and hence with an increase in testing rate more number of cases are reported as compared to other states. Kerala on the other hand attains a peak on the 25$^{th}$ of March, 2020, showing that the reduction of daily number of cases in Kerala cannot be attributed to the 21-day national lockdown. The Kerala state government adopted several stringent measure including a state lock down on the 23$^{rd}$ of March, 2020\cite{KL}, contact tracing and compulsory quarantine for all foreign travellers returning to the country\cite{KLmodel}; these early measures seem to have worked out for Kerala in flattening the curve.  \newline

Looking simply at the smoothened data does not give the entire picture. Even though the absolute numbers may increase for some states, it is not an indication that the lockdown is not working. So we parallelly consider the percent changes in the time series data of the daily confirmed cases. We see that in the case of Kerala, the percentage change in the number of daily cases drops drastically around the 25$ ^{th} $ day and then remain to be less than ~$2.5\% $. The percentage changes oscillate near 0 while constantly loosing amplitude, i.e. the magnitude of the percentage change, and then we see an abrupt rise. This steep increase in the percentage change towards the end of the lock down period can be attributed to the small number of daily confirmed cases which make a small change in absolute numbers to become relatively large. This success of the state of Kerala in reducing the absolute numbers as well as the relative changes can be attributed to their early interventions and prior experience on handling an epidemic\cite{Nippah}. In the case of Maharashtra, we observe that despite the absolute number of cases are increasing the percentage change in the daily confirmed cases reduce in magnitude approximately one week after the lock down peaking at the 14$ ^{th} $ day, then start oscillating around 0 towards the end of the lock down period. This suggests that the lock down has been successful in reducing the rate of growth of the pandemic in Maharashtra. A similar trend as Maharashtra is observed in Karnataka, Rajasthan, Madhya Pradesh and Uttar Pradesh. Punjab, Delhi and Tamil Nadu show a uniform trend in the percent changes of daily cases showing oscillating changes which have similar amplitudes which means that despite the lock down the number of confirmed cases are likely to grow in the coming days. In West Bengal we see that there is a spike a few days after the lockdown and percentage changes become small and closer to zeros near the end of the lock down period. There are huge gaps in the plot as the state has recorded 0 cases on some of the days which could be attributed to having the lowest testing rate in the country. Telangana shows two spikes, one near the beginning and one at the end, with small changes oscillating near zero in between which along with the time series plot shows that the daily confirmed cases spike towards the end of the data meaning that the lockdown has not had an impact in Telangana. In Jammu and Kashmir we see a steady increase in percentage changes and then a decrease. This shows that towards the end of the lockdown period Jammu and kashmir was able to reduce the percentage change in daily confirmed cases.    \newline

As we can see from the figures, tables and graphs each state is performing differently, this may be due to the way that the lock downs have been implemented in each of the states. While Maharashtra government declared a lock down on the 23$ ^{rd} $ of March and took early measures to reduce physical contact, like stopping the local trains and auto-rickshaws or reducing working staff in government by half or shutting down schools and colleges. Delhi, the state with second highest confirmed cases, despite early measures take by the state government Delhi saw several incidences where the lock down measures and protocol was not followed properly, like the one at the Nizamuddin Markeez which was attended by 3500 people\cite{Tablighi} or infection of a doctor in a Mohalla clinic which led to the quarantine of over 800 people\cite{mohallaclin}. Delhi also saw a huge number of migrant labourers fleeing from the capital due to a fear of lack of food during the 21-day lock down. West Bengal on the other hand has been conducting very few tests\cite{Testingrate} and even though they have take many health and administrative measures to tackle the pandemic, one cannot comment on the success or failure of West Bengal state government in curbing the virus due to its low testing rate. Rajasthan's major industry is the tourism with around 24,93,431 foreign tourists recorded in 2018-19\cite{RJtourist}, this high number of oversees tourist renders the state vulnerable to the pandemic. Bhilwara, in Rajasthan, emerged as a hotspot with 27 confirmed cases and and 2 deaths as of 3$^{rd}$ April and a lock down was imposed as soon as the first case was detected. The district collector rolled out a very aggressive plan for containment including isolation of the district, with its borders sealed, screening in the city and isolation of patients. A containment zone of 2kms was mapped around the residence of any infected individual and all those with foreign travel history were isolated and checked\cite{Bhilwara1}\cite{Bhilwara2}. \newline

\begin{table}[H]
	\centering
	\begin{tabular}{|l|l|l|l|}
		\hline
		Date           & Predicted & Lower limit  & Upper limit  \\ \hline
		April 18, 2020 & 14488     & 13970        & 15054        \\
		April 19, 2020 & 15601     & 14484        & 16816        \\
		April 20, 2020 & 16804     & 15050        & 18731        \\ 
		April 21, 2020 & 18103     & 15678        & 20786        \\
		April 22, 2020 & 19485     & 16352        & 22969        \\
		April 23, 2020 & 20928     & 17067        & 25254        \\
		April 24, 2020 & 22451     & 17827        & 27662         \\
		April 25, 2020 & 24059     & 18641        & 30195         \\
		April 26, 2020 & 25751     & 19514        & 32849         \\
		April 27, 2020 & 27528     & 20446        & 35625         \\
		April 28, 2020 & 29390     & 21273        & 38524         \\
		April 29, 2020 & 31339     & 22149        & 41545         \\
		April 30, 2020 & 33375     & 23076        & 44690          \\
		May 1, 2020    & 35501     & 24058        & 47961          \\
		May 2, 2020    &  37721    & 25098        & 51360          \\
		May 3, 2020    & 40033     & 26196        & 54887          \\
		May 4, 2020    & 42439     & 27352        & 58542          \\
		May 5, 2020    & 44940     & 28569        & 62328           \\
		May 6, 2020    &  47539    &   29850      & 66245           \\
		May 7, 2020    &   50237   &  31194       & 70296        	\\
		\hline                          
	\end{tabular}
	\label{Table:2}
	\caption{\small \sl Table of forecasted values, using ARIMA modelling for each state, of the total number of confirmed cases till May 7, 2020 with lower and upper limits obtained by rounding off the lower and upper prediction intervals of the forecast}
\end{table}

  \citeauthor{outbreakanal} et al. used data from the early-stages of the epidemic and an exponential model to predict that by the end of April, 2020 the number of cases in India would be 75,000\cite{outbreakanal}. Using ARIMA models we were able to forecast the daily cases for each state and hence find the cumulative cases and their $ 95\% $ prediction intervals. These estimates based on ARIMA model for each state provide a much better approximation for the total number of cases. On the 27$^{th}$ of March, India had 27,982 confirmed cases\cite{MOHFW} and the model predicts 27,528 cases and the doubling rate by the 7$ ^{th}$of may would be around 7 days.  \newline

\section{Conclusion}
We see that despite of a national strategy for containment, each state has performed differently and its performance depends upon the ground level implementation and the strategies adopted by the states themselves. This leads us to wonder whether there are other factors that could explain the success or failure of the lock down strategy. For example, the literacy rate of Kerala is 94$\%$ whereas that in Rajasthan is 66.11$ \% $ or that the population density is 11,320 people per $ km^{2} $ in Delhi while it is only 17 people per $ km^{2} $ in Arunachal Pradesh\cite{census}. Other parameters could be the number of hospital beds, doctors, police stations, expenditure on health and education infrastructure etc. This requires further investigation and could be one of the follow up studies done after this. \newline

The limitations of this study are:
\begin{enumerate}
	\item The data is dependent on the testing capacity and testing rate of each state and hence does not represent the true number of cases. 
	\item The models do not take into consideration arbitrary events like the one in Nizamuddin and similarly cannot account for other measures taken in the future by the government. For instance, with the lifting of the lock down on May 3$ ^{rd} $ the proposed model would not take this change into account and hence would need a modification.
	\item The predictions made for each state are only valid until the conditions that have been prevalent during the duration data collection remain the same.
\end{enumerate} 

What we learn from looking at the different states and their performance is that applying the same containment plain uniformly throughout the country is not very effective when compared to the locally planned strategies. For example, the policies of the state government of Kerala, Maharashtra and Rajasthan are much more effective at curbing the spread of the virus than Karnataka, Tamil Nadu and Telangana which are operating according to the national strategy. A uniform policy does not discriminate between the states based on the demographics of the population i.e. a large number of daily wage workers hail from the states of Uttar Pradesh and Bihar and cannot afford to stay at home for 21 days whereas Kerala, having a very high literacy rate and people working in professional sector, can afford to do so. \newline

The idea behind breaking down the strategies according to state and then district can be understood with this simple example. The Bhilwara model of stringent lockdown and ruthless containment cannot be applied to the Dharavi area of Mumbai which has a very high population density and any sort of containment could lead to the area becoming a hotspot due to lack of sanitation, medical facilities and education.\newline

In the future while developing policies, regarding epidemic management, we must do so in a manner that focusses on local containment rather than central containment.  The methodology applied here can be used to investigate the situation in other countries or regions as well

\section*{Declaration of Interests}
We declare no competing interests. 

\section{Acknowledgement}
I would like to thank Dr. Amiya Ranjan Bhowmick, Department of Mathematics, Institute of Chemical Technology, Mumbai, India, for his guidance and support throughout the project.

\printbibheading 
\printbibliography[type = book, title = {Books}]
\printbibliography[type = article, title = {Articles}]
\printbibliography[type = online, title = {Online Resources}]

\section{Appendix}
\subsection{Figures}
\begin{figure}[H]
	\centering
	\begin{subfigure}[b]{0.2\textwidth}
		\includegraphics[width=\textwidth]{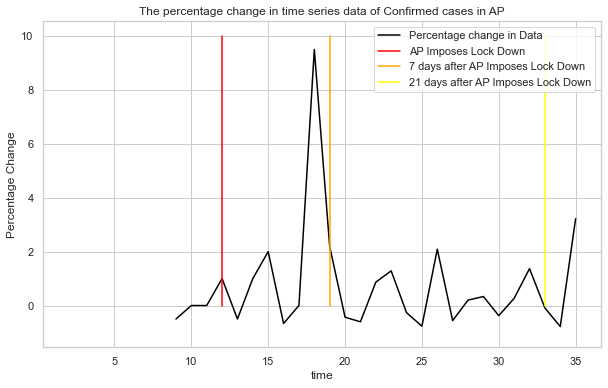}
	\end{subfigure}
	\begin{subfigure}[b]{0.2\textwidth}
		\includegraphics[width=\textwidth]{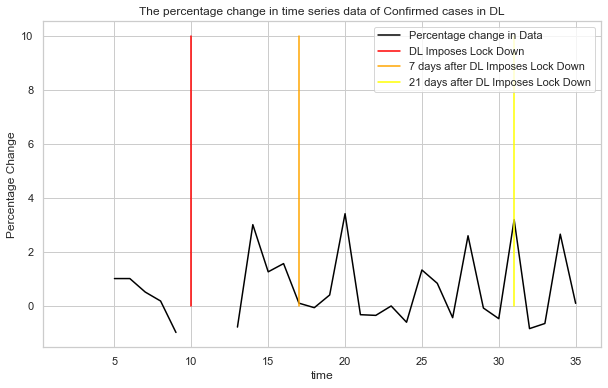}
	\end{subfigure}
	\begin{subfigure}[b]{0.2\textwidth}
		\includegraphics[width=\textwidth]{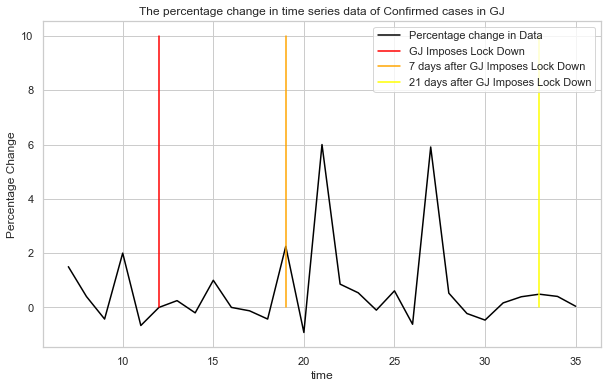}
	\end{subfigure}
	\begin{subfigure}[b]{0.2\textwidth}
		\includegraphics[width=\textwidth]{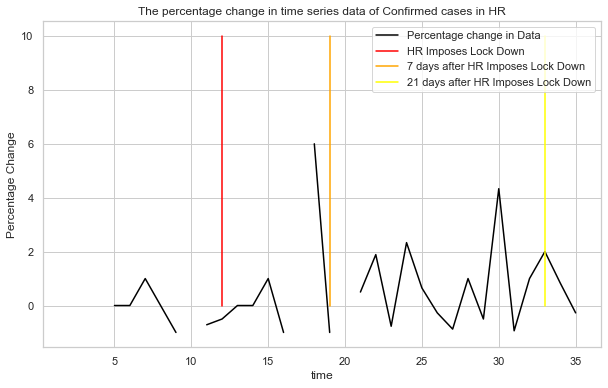}
	\end{subfigure}

	\begin{subfigure}[b]{0.2\textwidth}
		\includegraphics[width=\textwidth]{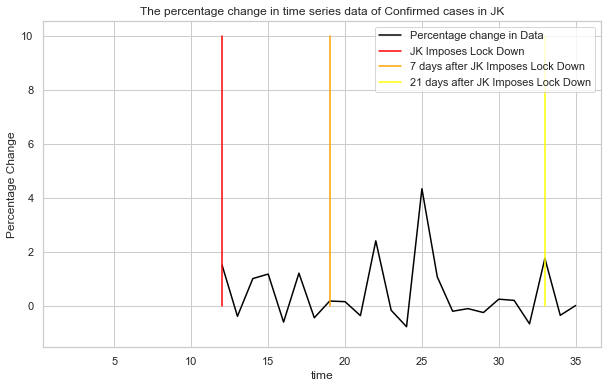}
	\end{subfigure}
	\begin{subfigure}[b]{0.2\textwidth}
		\includegraphics[width=\textwidth]{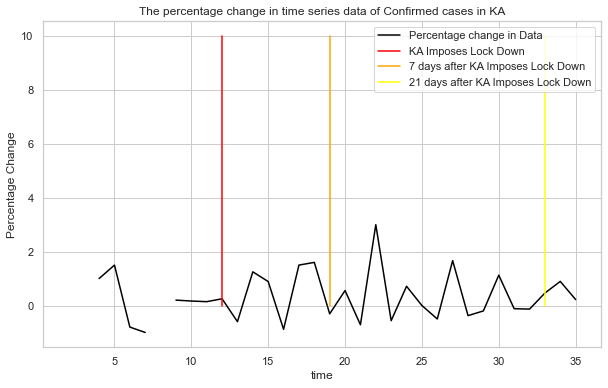}
	\end{subfigure}
	\begin{subfigure}[b]{0.2\textwidth}
		\includegraphics[width=\textwidth]{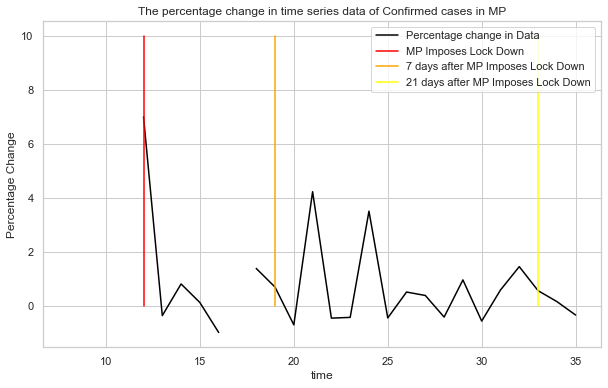}
	\end{subfigure}
	\begin{subfigure}[b]{0.2\textwidth}
		\includegraphics[width=\textwidth]{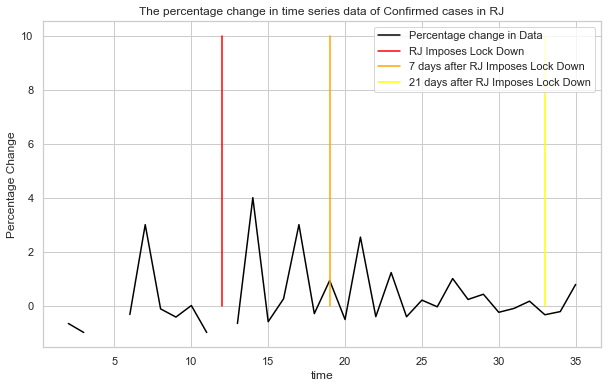}
	\end{subfigure}

	\begin{subfigure}[b]{0.2\textwidth}
		\includegraphics[width=\textwidth]{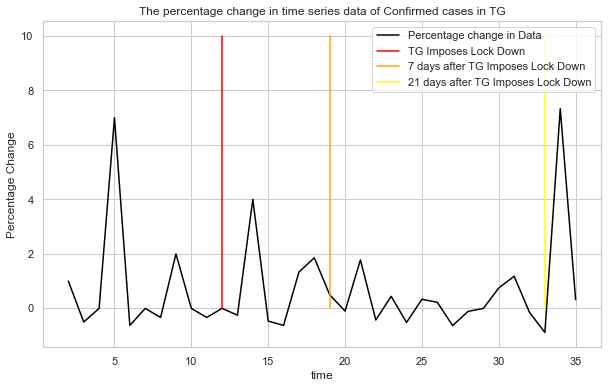}
	\end{subfigure}
	\begin{subfigure}[b]{0.2\textwidth}
		\includegraphics[width=\textwidth]{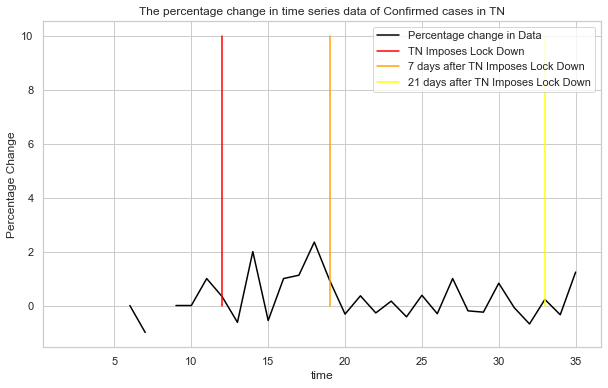}
	\end{subfigure}
	\begin{subfigure}[b]{0.2\textwidth}
		\includegraphics[width=\textwidth]{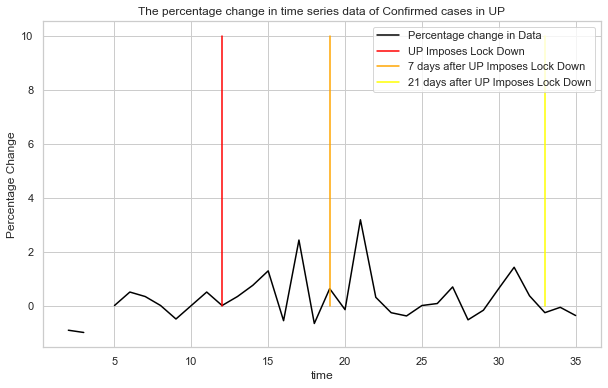}
	\end{subfigure}
	\begin{subfigure}[b]{0.2\textwidth}
		\includegraphics[width=\textwidth]{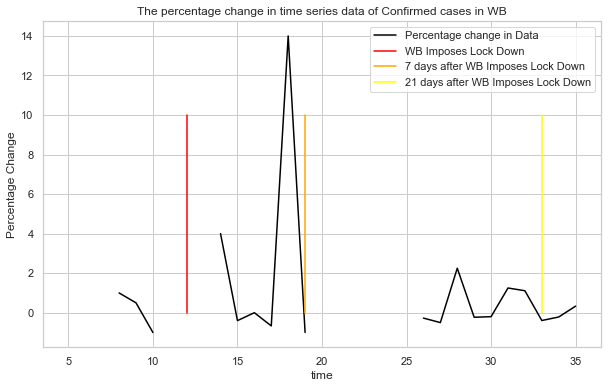}
	\end{subfigure}
\label{fig:4}
\caption{Percentage Changes in the daily number of cases in (from left to right) Andhra Pradesh, Delhi, Gujarat, Haryana, Jammu and Kashmir, Karnataka, Madhya Pradesh, Rajasthan, Telangana, Tamil Nadu, Uttar Pradesh, West Bengal}
\end{figure}
\begin{figure}[H]
	\centering
	\begin{subfigure}[b]{0.2\textwidth}
		\includegraphics[width=\textwidth]{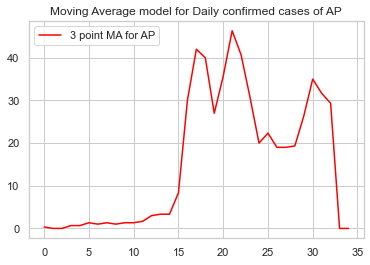}
	\end{subfigure}
	\begin{subfigure}[b]{0.2\textwidth}
		\includegraphics[width=\textwidth]{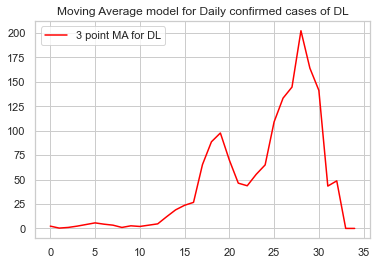}
	\end{subfigure}
	\begin{subfigure}[b]{0.2\textwidth}
		\includegraphics[width=\textwidth]{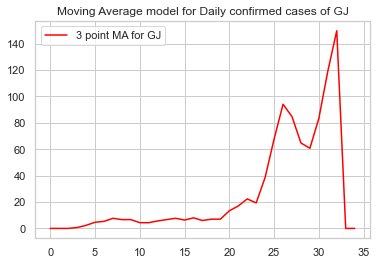}
	\end{subfigure}
	\begin{subfigure}[b]{0.2\textwidth}
		\includegraphics[width=\textwidth]{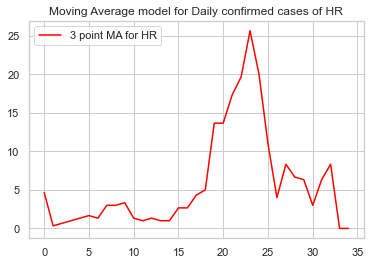}
	\end{subfigure}
	
	\begin{subfigure}[b]{0.2\textwidth}
		\includegraphics[width=\textwidth]{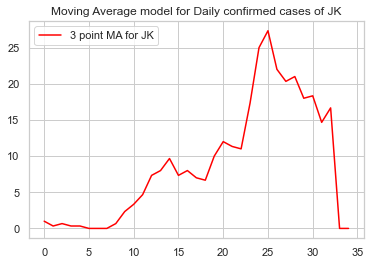}
	\end{subfigure}
	\begin{subfigure}[b]{0.2\textwidth}
		\includegraphics[width=\textwidth]{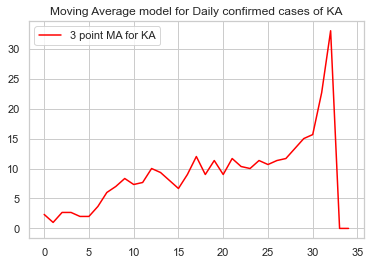}
	\end{subfigure}
	\begin{subfigure}[b]{0.2\textwidth}
		\includegraphics[width=\textwidth]{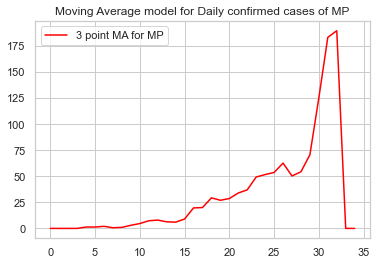}
	\end{subfigure}
	\begin{subfigure}[b]{0.2\textwidth}
		\includegraphics[width=\textwidth]{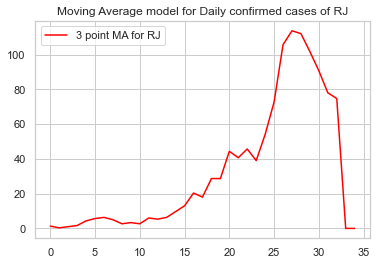}
	\end{subfigure}
	
	\begin{subfigure}[b]{0.2\textwidth}
		\includegraphics[width=\textwidth]{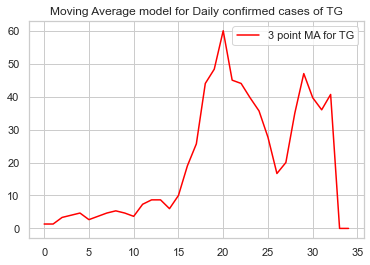}
	\end{subfigure}
	\begin{subfigure}[b]{0.2\textwidth}
		\includegraphics[width=\textwidth]{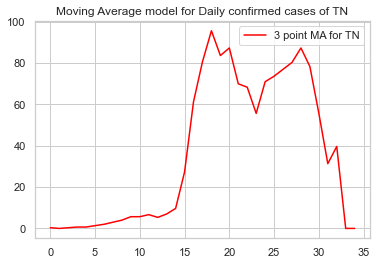}
	\end{subfigure}
	\begin{subfigure}[b]{0.2\textwidth}
		\includegraphics[width=\textwidth]{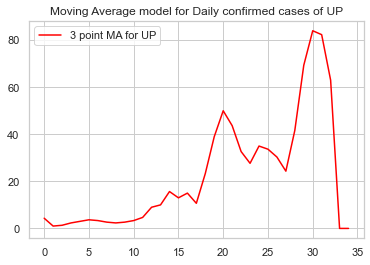}
	\end{subfigure}
	\begin{subfigure}[b]{0.2\textwidth}
		\includegraphics[width=\textwidth]{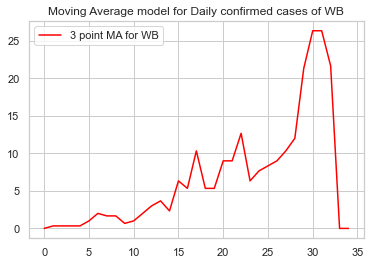}
	\end{subfigure}
	\label{fig:5}
	\caption{Moving averages of the daily number of cases in (from left to right) Andhra Pradesh, Delhi, Gujarat, Haryana, Jammu and Kashmir, Karnataka, Madhya Pradesh, Rajasthan, Telangana, Tamil Nadu, Uttar Pradesh, West Bengal}
\end{figure}

\begin{figure}[H]
	\centering
	\begin{subfigure}[b]{0.2\textwidth}
		\includegraphics[width=\textwidth]{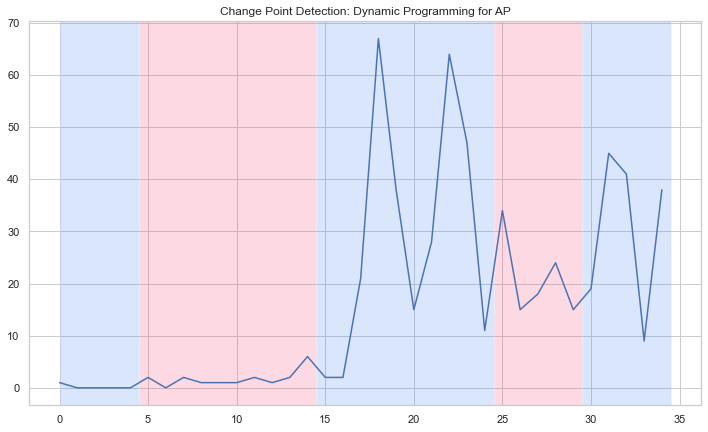}
	\end{subfigure}
	\begin{subfigure}[b]{0.2\textwidth}
		\includegraphics[width=\textwidth]{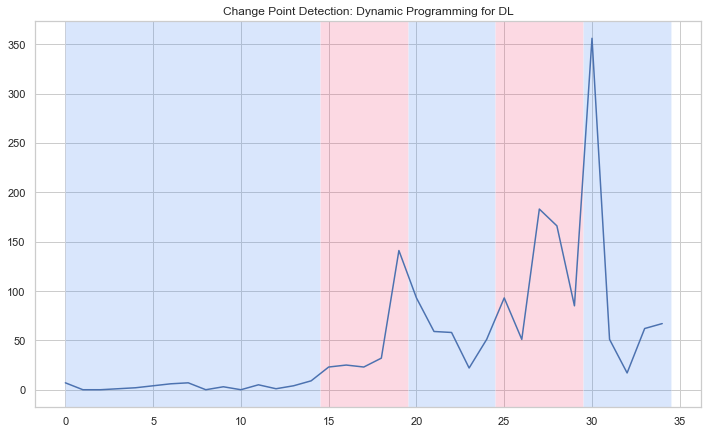}
	\end{subfigure}
	\begin{subfigure}[b]{0.2\textwidth}
		\includegraphics[width=\textwidth]{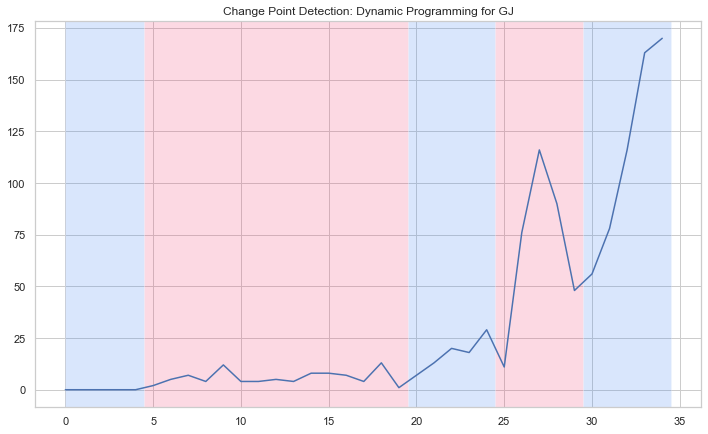}
	\end{subfigure}
	\begin{subfigure}[b]{0.2\textwidth}
		\includegraphics[width=\textwidth]{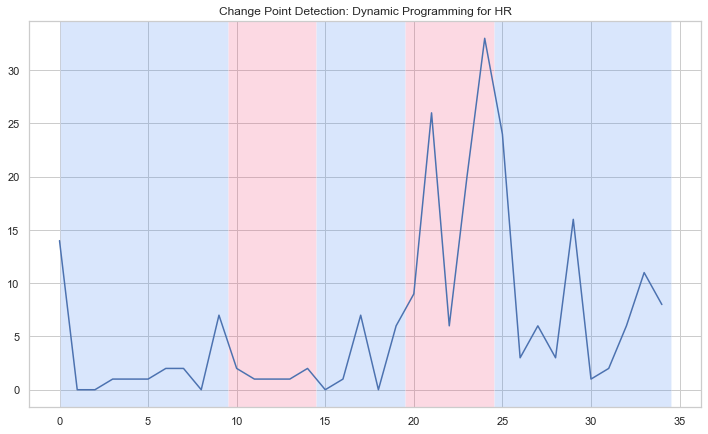}
	\end{subfigure}
	
	\begin{subfigure}[b]{0.2\textwidth}
		\includegraphics[width=\textwidth]{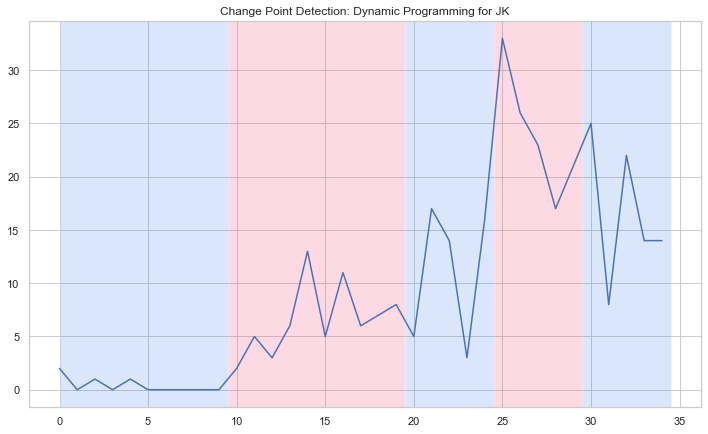}
	\end{subfigure}
	\begin{subfigure}[b]{0.2\textwidth}
		\includegraphics[width=\textwidth]{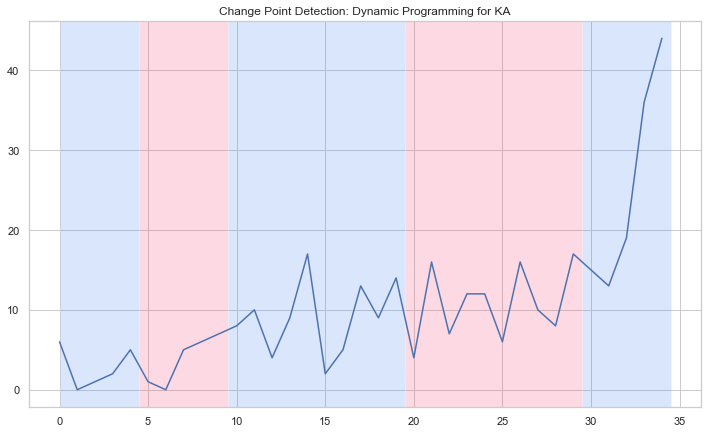}
	\end{subfigure}
	\begin{subfigure}[b]{0.2\textwidth}
		\includegraphics[width=\textwidth]{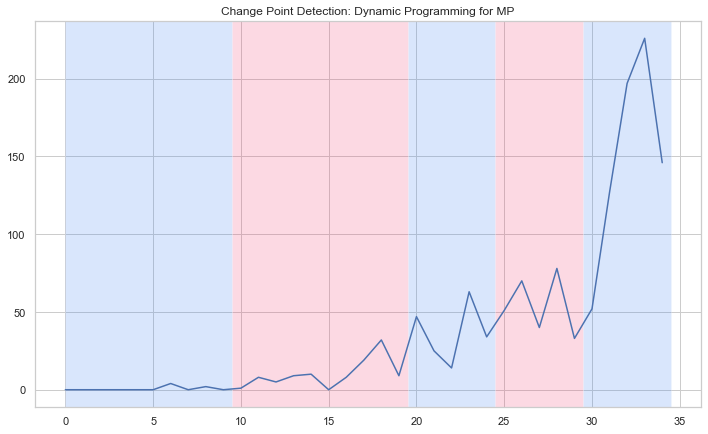}
	\end{subfigure}
	\begin{subfigure}[b]{0.2\textwidth}
		\includegraphics[width=\textwidth]{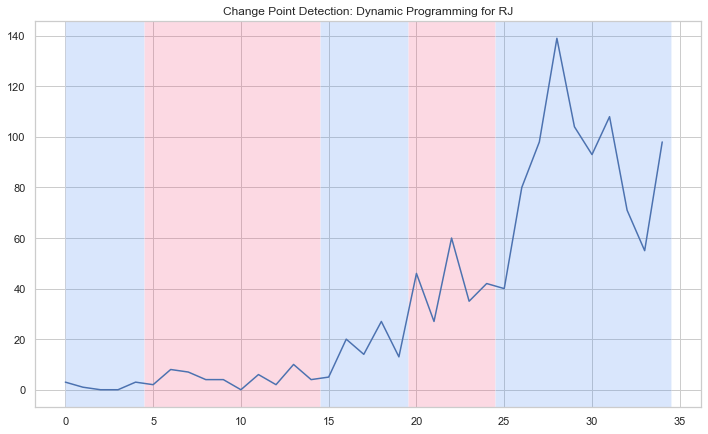}
	\end{subfigure}
	
	\begin{subfigure}[b]{0.2\textwidth}
		\includegraphics[width=\textwidth]{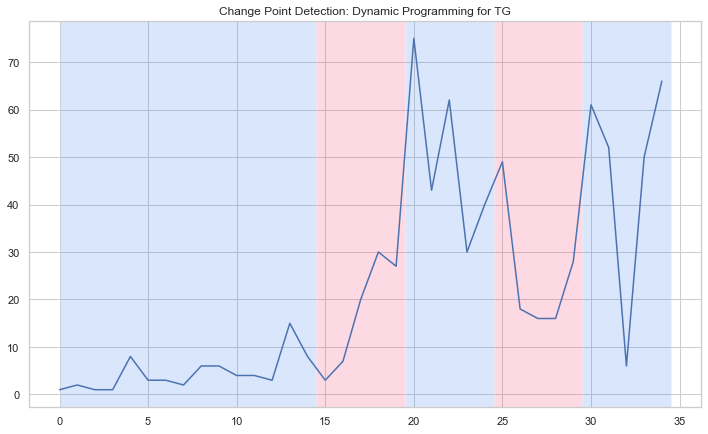}
	\end{subfigure}
	\begin{subfigure}[b]{0.2\textwidth}
		\includegraphics[width=\textwidth]{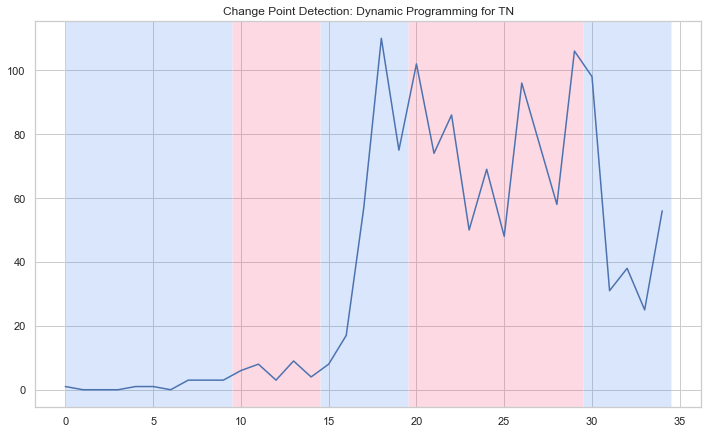}
	\end{subfigure}
	\begin{subfigure}[b]{0.2\textwidth}
		\includegraphics[width=\textwidth]{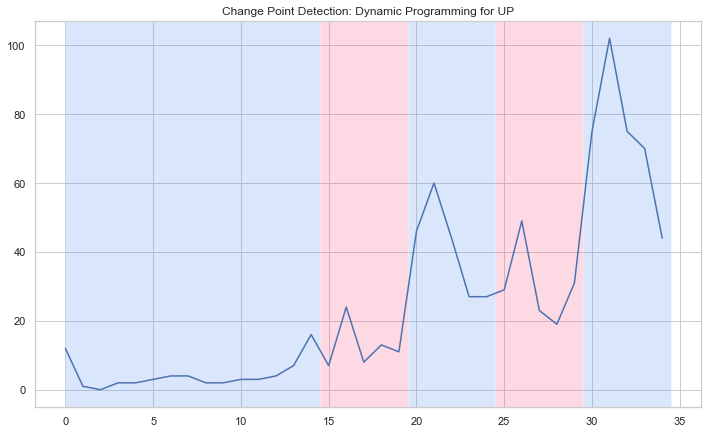}
	\end{subfigure}
	\begin{subfigure}[b]{0.2\textwidth}
		\includegraphics[width=\textwidth]{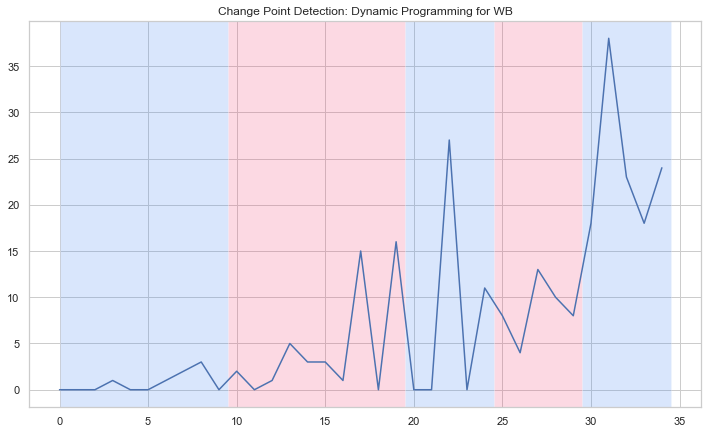}
	\end{subfigure}
	\label{fig:6}
	\caption{Change Points of the daily number of cases in (from left to right) Andhra Pradesh, Delhi, Gujarat, Haryana, Jammu and Kashmir, Karnataka, Madhya Pradesh, Rajasthan, Telangana, Tamil Nadu, Uttar Pradesh, West Bengal}
\end{figure}

\begin{figure}[H]
	\centering
	\begin{subfigure}[b]{0.2\textwidth}
		\includegraphics[width=\textwidth]{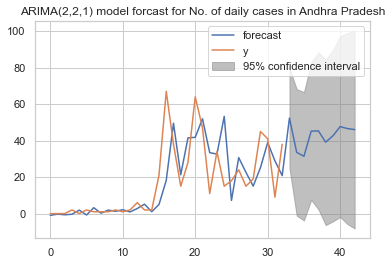}
	\end{subfigure}
	\begin{subfigure}[b]{0.2\textwidth}
		\includegraphics[width=\textwidth]{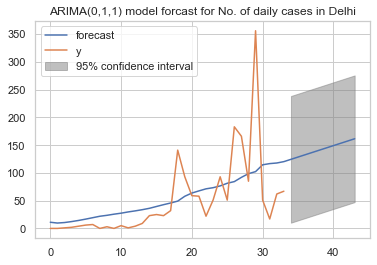}
	\end{subfigure}
	\begin{subfigure}[b]{0.2\textwidth}
		\includegraphics[width=\textwidth]{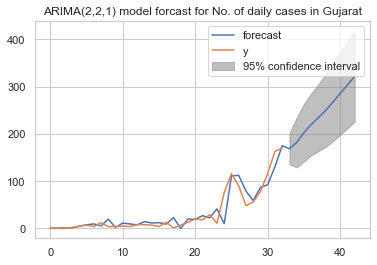}
	\end{subfigure}
	\begin{subfigure}[b]{0.2\textwidth}
		\includegraphics[width=\textwidth]{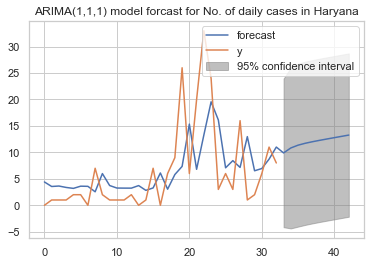}
	\end{subfigure}
	
	\begin{subfigure}[b]{0.2\textwidth}
		\includegraphics[width=\textwidth]{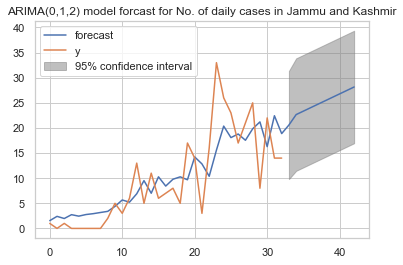}
	\end{subfigure}
	\begin{subfigure}[b]{0.2\textwidth}
		\includegraphics[width=\textwidth]{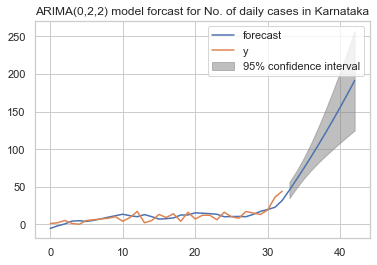}
	\end{subfigure}
	\begin{subfigure}[b]{0.2\textwidth}
		\includegraphics[width=\textwidth]{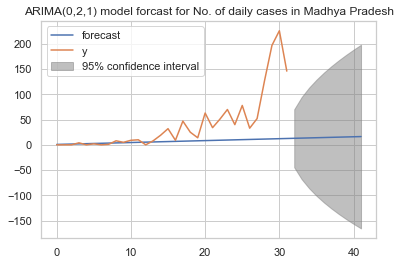}
	\end{subfigure}
	\begin{subfigure}[b]{0.2\textwidth}
		\includegraphics[width=\textwidth]{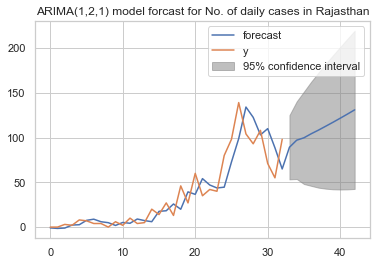}
	\end{subfigure}
	
	\begin{subfigure}[b]{0.2\textwidth}
		\includegraphics[width=\textwidth]{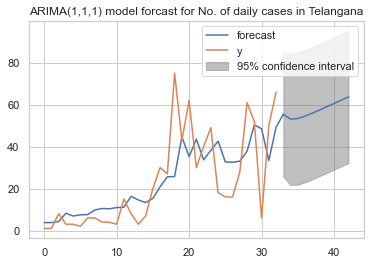}
	\end{subfigure}
	\begin{subfigure}[b]{0.2\textwidth}
		\includegraphics[width=\textwidth]{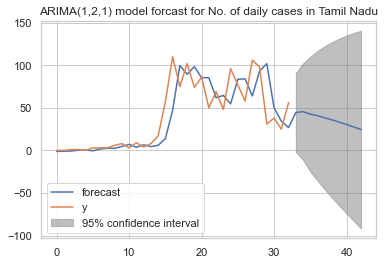}
	\end{subfigure}
	\begin{subfigure}[b]{0.2\textwidth}
		\includegraphics[width=\textwidth]{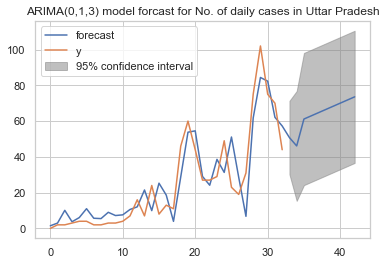}
	\end{subfigure}
	\begin{subfigure}[b]{0.2\textwidth}
		\includegraphics[width=\textwidth]{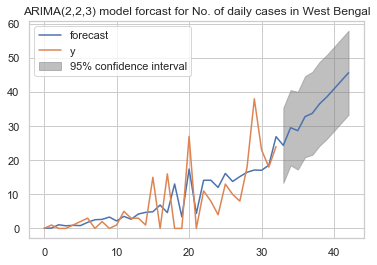}
	\end{subfigure}
	\label{fig:7}
	\caption{ARIMA models of the daily number of cases in (from left to right) Andhra Pradesh, Delhi, Gujarat, Haryana, Jammu and Kashmir, Karnataka, Madhya Pradesh, Rajasthan, Telangana, Tamil Nadu, Uttar Pradesh, West Bengal}
\end{figure}

\end{document}